\documentclass[oldversion]{aa}
\usepackage{graphicx}
\usepackage{txfonts}
\usepackage{natbib}
\bibpunct{(}{)}{;}{a}{}{,} 

\begin{document}

   \title{Molecular clouds in the center of M81}


   \author{V. Casasola
          \inst{1,2} \and
          F. Combes\inst{2} \and
          D. Bettoni \inst{3}
          \and
          G. Galletta \inst {1}
          }

   \offprints{V. Casasola}

   \institute{Dipartimento di Astronomia, Universit\`a di Padova, Vicolo
               dell'Osservatorio 2, I-35122, Padova\\
              \email{viviana.casasola@unipd.it, giuseppe.galletta@unipd.it}
         \and
             LERMA, Observatoire de Paris, 61 Avenue de l'Observatoire, F-75014 Paris\\
             \email{francoise.combes@obspm.fr}
         \and
             INAF, Osservatorio Astronomico di Padova, Vicolo 
             dell'Osservatorio 5, I-35122, Padova\\
                \email{daniela.bettoni@oapd.inaf.it}
             }

   \date{Received , ; Accepted , }


  \abstract
  {We investigate the molecular gas content and the excitation and fragmentation properties in the central region of the spiral
   galaxy Messier 81 in both the $^{12}$CO(1-0) and $^{12}$CO(2-1) transitions. 
   We have recently observed the two transitions of CO in the M~81 center with A, B, and HERA receivers 
   of the IRAM 30-m telescope. We find no CO emission in the
   inner $\sim$ 300 pc and a weak molecular gas clump structure at a distance of around 460 pc
   from the nucleus. Observations of the first two CO transitions allowed us to compute the line ratio, and 
   the average $I_{21}/I_{10}$ ratio is 0.68 for the M~81 center. This low value, atypical
   both of the galactic nuclei of spiral galaxies and of interacting systems, is probably associated to diffuse
   gas with molecular hydrogen density that is not high enough to excite the CO molecules. After analyzing the
   clumping properties of the molecular gas in detail, we identify very massive giant molecular associations (GMAs) 
   in CO(2-1) emission with masses of $\sim$ 10$^{5}$ M$_\odot$ and diameters of $\sim$ 250 pc. The deduced 
   $N(H_{2})/I_{CO}$ ratio for the individually resolved GMAs, assumed to be virialized, is a factor of 
   $\sim$ 15 higher than the \textit{standard} Galactic value, showing - as
   suspected - that the X ratio departs significantly from the mean for galaxies with an unusual physics of the molecular gas.}


   \maketitle

%

\section{Introduction}

   The molecular gas content of Messier 81, one of the nearest face-on spirals 
 \citep[SA(s)ab;][]{de vaucouleurs}, has always been a puzzle  as seen in the first 
  observations by \citet{solomon} and \citet{combes1}. The CO emission
  appears very weak in this galaxy, especially in the central regions, and the molecular content
seems to be confined to the HII regions in the spiral arms, between 4 and 7 kpc 
from the center \citep{solomon2,brouillet1,brouillet2,sage,sakamoto2}. 
Two studies \citep{solomon,combes1} searched the molecular gas in the center 
of M~81 but only found upper limits or  failed to detect any - probably due to the low CO emission - 
so they did not draw any conclusions from this result. The first detection of the molecular gas in the
center of M~81 was by \citet{sage}, who found that the peak intensities are a factor 4 lower 
than those observed by \citet{brouillet1} in the outer disk. After the weak detection by \citet{sage}, 
only \citet{sakamoto2} have studied the CO emission
in the center of M~81 and also found a particular molecular gas physics.
M~81 has also been studied in all components of its interstellar medium (ISM), from radio
\citep[e.g.][]{beck,bietenholz}, through optical and UV \citep[e.g.][]{ho},
to X-ray bands \citep[e.g.][]{pellegrini}. 

  In addition to the weak CO emission, the HI profile also reflects a central deficiency in 
  atomic gas. The HI gas is abundant only in the inner spiral arms of the galaxy \citep{visser,allen2}, 
  and the shapes and the motions observed in HI in the spiral arms quite clearly obey the 
  predictions of the density-wave theory. This strong density wave is attributed to the tidal 
  interaction with the other galaxies of the M~81 group, especially M 82 and NGC 3077 \citep{kaufman}.

   In general, interacting galaxies like M~81 possess higher CO luminosity by almost an order of
magnitude than non-interacting galaxies \citep{braine2,combes2,casasola}. Then, M~81 seems 
an exception remarkably poor for its CO emission. For this reason, M~81 is a kind of
prototype of CO-poor galaxies, as is Andromeda or the center of NGC 7331, a lesser extent.

M~81 is also a good candidate for exploring the problem of the varying CO to H$_2$ conversion ratio 
(X = N(H$_2$)/I$_{CO}$). This ratio is well-known for varying substantially in dwarf galaxies, for which 
the metallicity is deficient \citep{rubio,taylor}, while for galaxies like M~81 that are giants with 
normal metallicity, the problem of the low CO emission has not been yet explored much. 
The CO to H$_2$ conversion ratio for M~81 is suspected - on the base of the observations that are 
already available - of departing significantly from the mean \citep[$X=2.3 \times 10^{20}$ 
mol cm$^{-2}$ (K km s$^{-1}$)$^{-1}$,][]{strong}, because galaxies with low CO content, 
such as M~31, NGC~55, or LMC, show a much larger $X$ factor \citep[][]{nakai}.

   In this paper, we assume M~81 galaxy at a distance of 3.6 Mpc and adopt 
   $H_{0}$ = 70 km s$^{-1}$ Mpc$^{-1}$.
   We present the CO(1-0) and CO(2-1) observations of the central $\sim1.5$ kpc 
   of M~81 carried out at the IRAM 30-m telescope. The primary goal of this work is to investigate 
   the amount and distribution of the molecular gas component in the central regions. 

   The structure of the paper is as follows. In Sect. 2 the observations performed and the
 techniques of the data reduction used are described. In Sect. 3 we present the observational results,  
 and in particular, the $R_{21}=I_{21}/I_{21}$ line ratio is studied and analyzed for all 
the offsets observed that give information on the physical properties of the gas in the M~81 center.
The clumpy nature of the molecular gas, the virial equilibrium, and the X-ratio of the resolved 
individual clouds are discussed in Sect. 4. Our results are compared with previous studies of 
the molecular gas in M~81 and discussed, giving their physical interpretation in Sect. 5.
Finally, in Sect. 6 we summarize the main results.


\section{Observations and data reduction}

%
   The observations were made from 2 from 6 January 2006 and from 31 March to
1 April 2006 at the IRAM 30-m telescope at Pico Veleta, near Granada (Spain).
We used two different configurations for the two observing runs
and we covered the central $90^{\prime\prime} \times 90^{\prime\prime}$, 
where at the distance of M~81, 1$^{\prime\prime}$ = 17.4 pc. The receiver-cabin 
optic system of the IRAM 30-m telescope is optimized to observe
with up to four different receivers simultaneously \citep{wild}.

   In the first observing run we used the A100, A230, B100, and B230 heterodyne
receivers of the telescope, which simultaneously observed at both 115 GHz (A100 and B100,
at $\lambda \simeq$ 2.6 mm, where $\lambda$ is the wavelength) and 230 GHz (A230
and B230, at $\lambda \simeq$ 1.3 mm). We used the 1 MHz back-ends with an effective
total bandwidth of 512 MHz at 2.6 mm and an autocorrelator at 1.3 mm. We also used the
4 MHz filterbanks with an effective total bandwidth of 1024 MHz at 1.3 mm.
   These arrangements provided a velocity coverage of 1300 km s$^{-1}$ in both the CO(1-0) 
and CO(2-1) lines, and all the measurements were performed in ``wobbler-switching'' mode. 
This observing mode has a  minimum phase time for spectral line observations of 2 sec 
and a maximum beam throw of $240^{\prime\prime}$. The advantage of the ``wobbler-switching'' 
mode is to give very flat baselines without any ripple in most cases. The half power beam 
widths (HPBW) are $22^{\prime\prime}$ and $12^{\prime\prime}$ for the CO(1-0) and CO(2-1) 
lines, respectively, and the typical system temperatures were $\sim$100 K at 115 GHz and
$\sim$400 K at 230 GHz. The positions observed with A and B receivers were 81 for both lines.

\begin{table}
   \caption[]{Fundamental parameters for M~81 (at a distance of 3.6 Mpc adopted for M~81,
   $1^{\prime\prime}$ is $\sim$17.4 pc).}
   \begin{center}
   \begin{tabular}{lcccc}
   \hline
   \hline
     R.A.  & Dec & V & $\nu$ & line \\
   (J2000.0)   & (J2000.0)     & (km/s)& GHz &\\
   \hline
   09:55:33 & 69:03:55 & -34 & 115.271 & $^{12}$CO(1-0)\\
   &             &     &230.538 & $^{12}$CO(2-1)\\
   \hline
   \end{tabular}
   \label{table1}
   \end{center}
   \end{table}

\begin{figure}
   \centering
   \includegraphics[width=7cm]{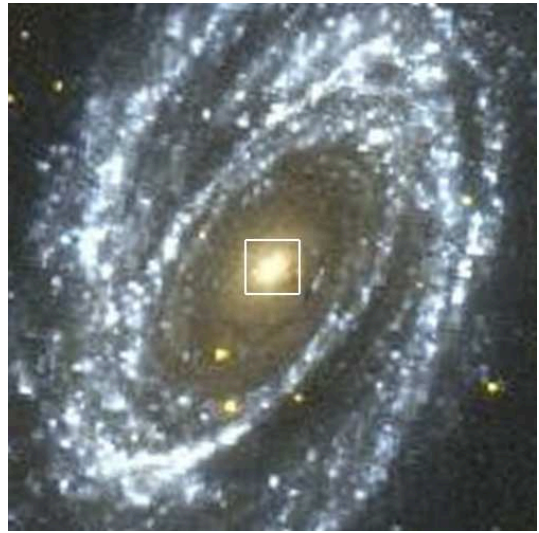}
   \caption{The field of our observations (white box, 40 arcsec in size) superposed on a GALEX 
composite image of M~81. In the figure, FUV is in the blue channel and NUV the yellow one.}
  \label{galex}
  \end{figure}

   In the second observing run, we used the Heterodyne Receiver Array HERA
\citep{schuster}, a focal array of 18 SIS receivers, 9 for each polarization,
tuned to the CO(2-1) line for M~81. The 9 pixels are arranged in the
form of a center-filled square and are separated by $24^{\prime\prime}$. The
sampling was $6^{\prime\prime}$, and a homogeneous mapping procedure was used to
regularly sweep a $12 \times 12$ pixel map, filling the intrinsic square of
$66^{\prime\prime} \times 66^{\prime\prime}$. The typical system temperature was
$\sim$400 K. Also in this case we used the ``wobbler-switching'' observing mode, and
the pointing accuracy was $\sim3^{\prime\prime}$ rms. The WILMA backend was used,
providing a 1 GHz bandwidth for each of the 18 receivers. The bands contain 2$\times$465 MHz (=930 MHz)
channels spaced by 2 MHz of resolution. The total bandwidth corresponds to 1300 km s$^{-1}$
with a velocity resolution of 2.6 km s$^{-1}$. The map realized with HERA included 144 positions.

\begin{figure*}
\centering
\includegraphics[height=17cm,angle=-90]{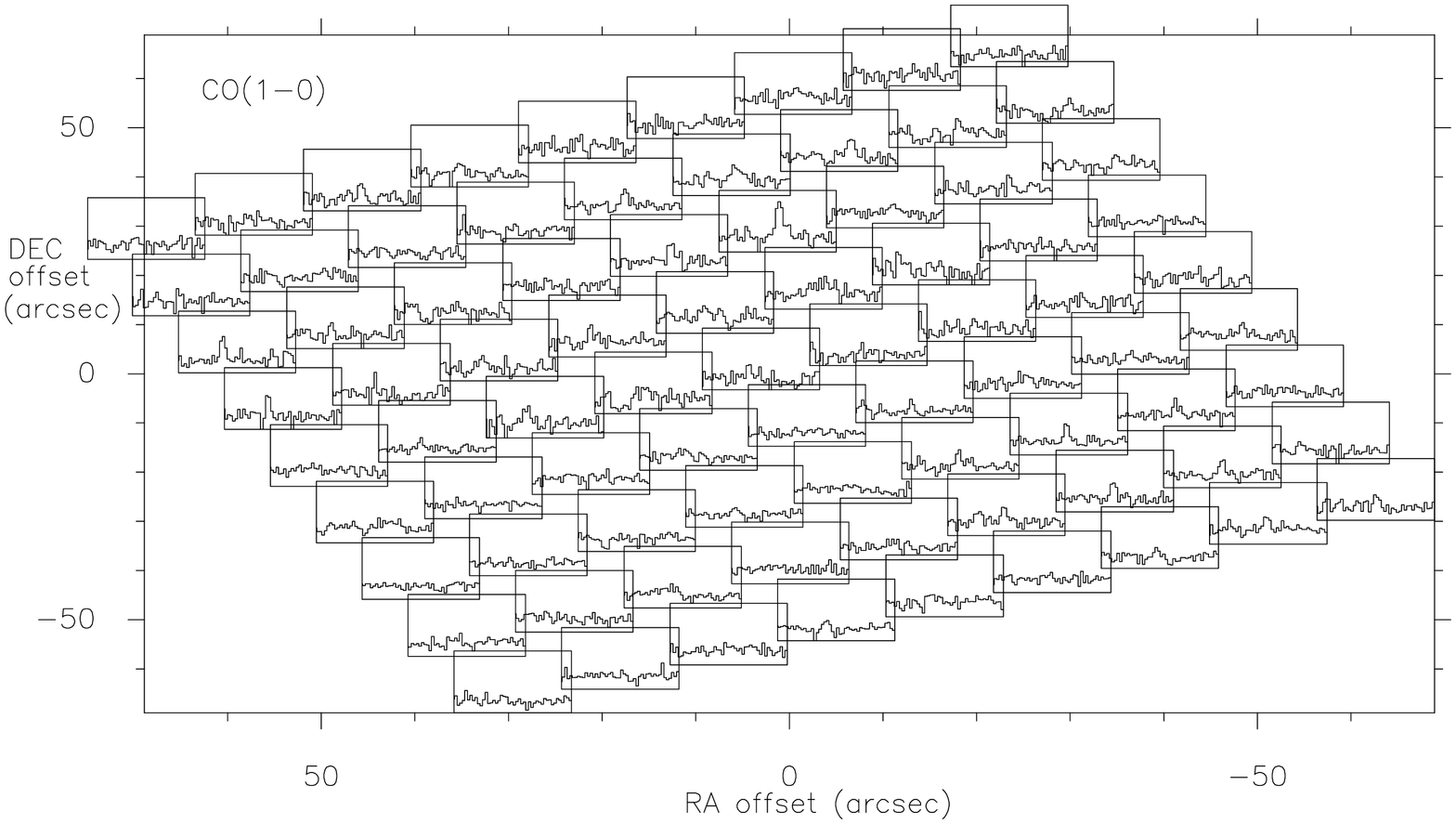}
\includegraphics[height=17cm,angle=-90]{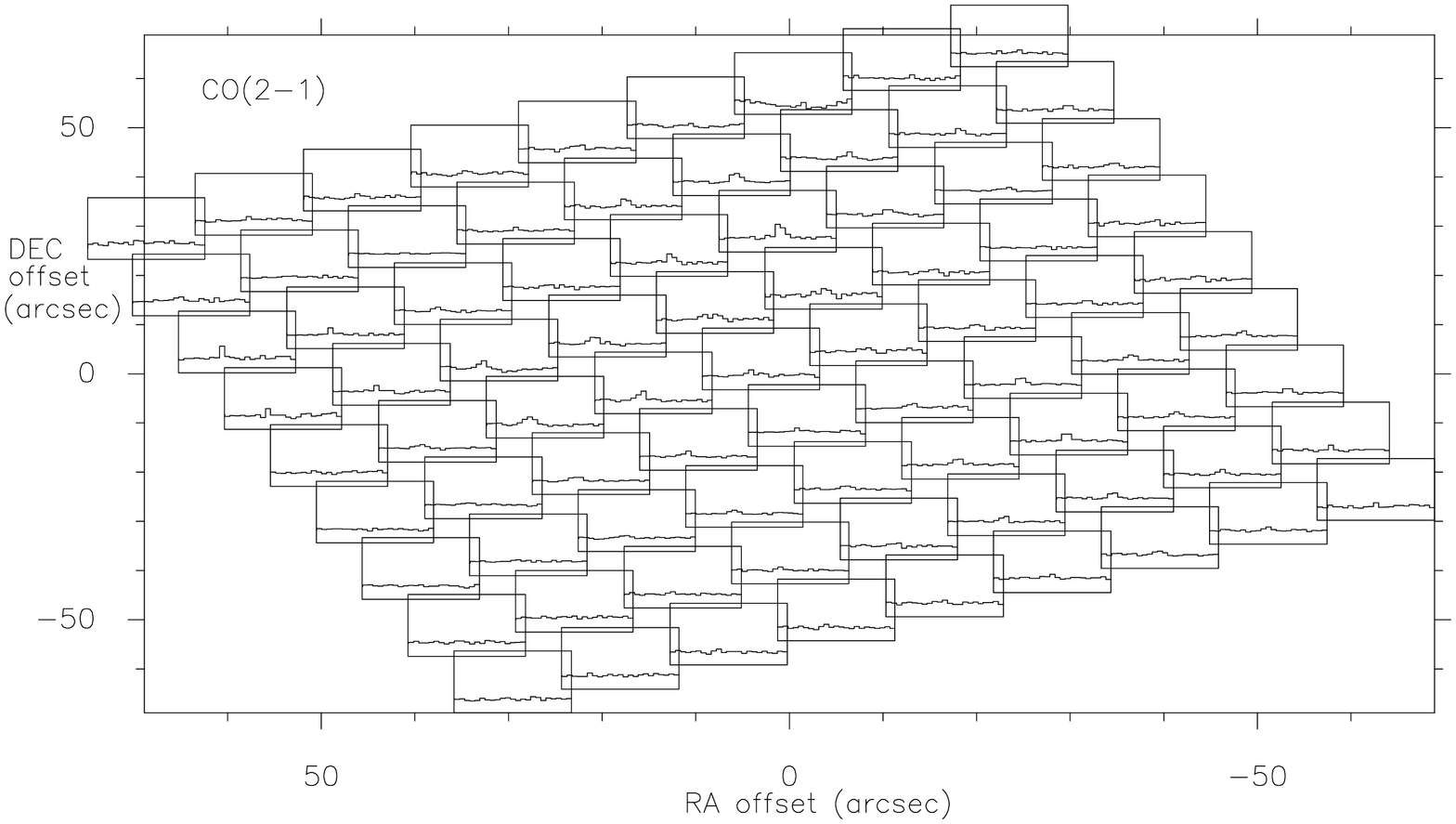}
\caption{Map of the observed M~81 central region. Each spectrum has a velocity
scale from -600 to 600 km s$^{-1}$, and a temperature scale from -20
to 70 mK. The positions are offsets relative to the 
M~81 nucleus assumed of coordinates RA$_{J2000.0}$ = 09$^{h}$ 55$^{m}$ 33$^{s}$, 
DEC$_{J2000.0}$ = 69$^{\circ}$ 03$^{\prime}$ 55$^{\prime\prime}$. 
\textit{Upper panel}: Observations of the $^{12}$CO(1-0) line. \textit{Bottom panel}:
Observations of the $^{12}$CO(2-1) line convolved to the $^{12}$CO(1-0) beam resolution 
($22^{\prime\prime}$).}
\label{fig1}
\end{figure*}

   The line intensity scale throughout this paper is in units of $T_{mb}$, the beam-average 
radiation temperature. The value of $T_{mb}$ is related to $T^{*}_{A}$, the equivalent antenna temperature 
-corrected  for rear spillover and ohmic losses- reported above the atmosphere,
by $\eta=T^{*}_{A}/T_{mb}$, where $\eta$ is the telescope main-beam efficiency, $\eta$  
defined as the ratio between the beam efficiency $B_{eff}$ and the forward efficiency $F_{eff}$, 
$\eta=B_{eff}/F_{eff}$. The beam efficiency is usually derived from continuum measurements of 
planets, while the forward efficiency is  obtained by pointing the antenna at different elevations 
and measuring the $T^{*}_{A}$ of the sky at each step. The sky temperatures are successively 
fitted by an exponential function of the air mass, and the forward efficiency is derived with the 
atmospheric opacity. At 115 GHz $\eta=0.79$, while $\eta_{B_{eff}}=0.57$ at 230 GHz.

   The data were reduced with the GILDAS \footnote{http://www.iram.fr/IRAMFR/GILDAS/} software
package. Some spectra with random, highly non-linear baselines were suppressed, and linear 
baselines were subtracted from all the remaining spectra. In Table \ref{table1} the fundamental 
parameters for the M~81 observations are summarized.

\section{Molecular gas emission}

   The region observed with the A and B receivers of the IRAM 30-m telescope in the two CO lines 
covers the central $\sim$ 1.6 kpc of M~81. The size and the location of the region covered by our 
CO observations are shown in Fig. \ref{galex} as a square superposed on a GALEX image of the galaxy.
The CO emission, both in CO(1-0) and CO(2-1) lines, 
comes from different regions of the galaxy nucleus. In agreement with \citet{sage} and 
\citet{sakamoto2}, we find that the nucleus presents some regions devoid of molecular gas,
but also that there are other regions in which the CO is clearly detected and the signal is strong 
(Fig. \ref{fig1}).

\begin{table*}[htbp]
   \centering
   \caption[]{CO(\textit{J}=1-0) and CO(\textit{J}=2-1) intensity lines where CO
   emission is detected (detection level $\geq2\sigma$).}
   \begin{tabular}{rrccccccc}
   \hline
   Offset                               &  Offset                            &$T_{mb}(1-0)$ & $\Delta$v(1-0)&   $\int{T_{mb}(1-0)}$dv &$T_{mb}(2-1)$  & $\Delta$v(2-1)& $\int{T_{mb}(2-1)}$dv  & $R_{21}$ \\
   $\Delta\alpha$ ($^{\prime\prime}$)   &  $\Delta\beta$ ($^{\prime\prime}$) &[mK]&[km s$^{-1}$]& [K km s$^{-1}$] &[mK]& [km s$^{-1}$]& [K km s$^{-1}$]        &          \\
   (1)&(2)&(3)&(4)&(5)&(6)&(7)&(8)&(9) \\
   \hline
-51.9	&	-30.1	&	29.04	&	74.01	&	2.15	&	10.23	&	83.21	&	0.85	&	0.40	\\
-47.2	&	-19.1	&	30.16	&	59.01	&	1.78	&	12.73	&	82.55	&	1.05	&	0.59	\\
-42.5	&	-8.0	&	28.34	&	88.23	&	2.50	&	14.55	&	115.60	&	1.68	&	0.67	\\
-37.8	&	3.0	&	16.97	&	79.81	&	1.35	&		&		&		&		\\
-36.2	&	-23.8	&	34.22	&	59.89	&	2.05	&	16.67	&	87.36	&	1.46	&	0.71	\\
-31.5	&	-12.7	&	38.18	&	52.72	&	2.01	&	23.87	&	73.12	&	1.75	&	0.87	\\
-26.8	&	-1.7	&	18.09	&	150.30	&	2.72	&	20.39	&	97.33	&	1.98	&	0.73	\\
-25.1	&	-28.4	&	31.67	&	61.07	&	1.93	&	25.13	&	52.02	&	1.31	&	0.68	\\
-20.4	&	-17.4	&	33.16	&	47.82	&	1.59	&		&		&		&		\\
-19.1	&	47.2	&	31.85	&	56.16	&	1.79	&		&		&		&		\\
-8.0	&	42.5	&	38.36	&	96.15	&	3.69	&	16.16	&	70.65	&	1.14	&	0.31	\\
-1.7	&	26.8	&	68.42	&	102.50	&	7.01	&	40.16	&	102.10	&	4.10	&	0.58	\\
3.0	&	37.8	&	27.99	&	152.80	&	4.28	&	24.86	&	116.50	&	2.90	&	0.68	\\
6.4	&	-15.7	&	12.52	&	113.40	&	1.42	&		&		&		&		\\
9.4	&	22.1	&	39.60	&	78.91	&	3.12	&		&		&		&		\\
14.1	&	33.1	&	34.47	&	66.19	&	2.28	&	22.30	&	80.29	&	1.79	&	0.78	\\
15.7	&	6.4	&	41.16	&	44.16	&	1.82	&	18.33	&	94.03	&	1.72	&	0.95	\\
22.1	&	-9.4	&	31.75	&	96.58	&	3.07	&	20.61	&	121.10	&	2.50	&	0.81	\\
26.8	&	1.7	&	21.40	&	67.65	&	1.45	&	22.95	&	58.55	&	1.34	&	0.93	\\
33.1	&	-14.1	&	21.76	&	43.95	&	0.96	&		&		&		&		\\
48.9	&	-7.7	&	58.60	&	32.25	&	1.89	&	19.20	&	52.02	&	1.00	&	0.53	\\
53.6	&	3.3	&	47.86	&	55.29	&	2.65	&	35.44	&	61.72	&	2.19	&	0.83	\\
\hline
    \end{tabular}
    \label{table2}
    \end{table*}

\begin{figure}[ht]
    \centering
		\includegraphics[width=9cm]{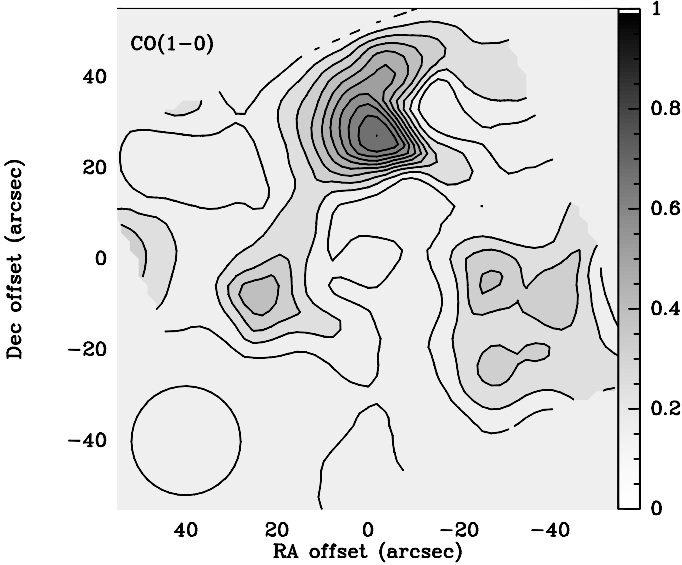}
	  \caption{Integrated $^{12}$CO(1-0) emission. The contour levels are between 0.701 to
     7.01 K km/s in steps of 0.701 K km/s. The beam of $22^{\prime\prime}$
     is indicated at the bottom left.}
	  \label{fig2}
   \end{figure}

   In particular, the central $\sim$300 pc region
   is devoid of CO(1-0) emission in agreement, for instance, with \citet{brouillet1} and \citet{sakamoto2}, 
   but there is a main molecular gas 
clump structure at a distance of around 460 pc from the nucleus in the northeast direction 
(Fig. \ref{fig2}), which is very similar to the maximum intensity detected by \citet{sakamoto2}. 
In addition to the CO arc in the northeast direction, we also detected some emission to the 
southwest one.

 \subsection{HERA-receiver results}
\begin{figure*}
   \centering
   \includegraphics[height=17cm,angle=-90]{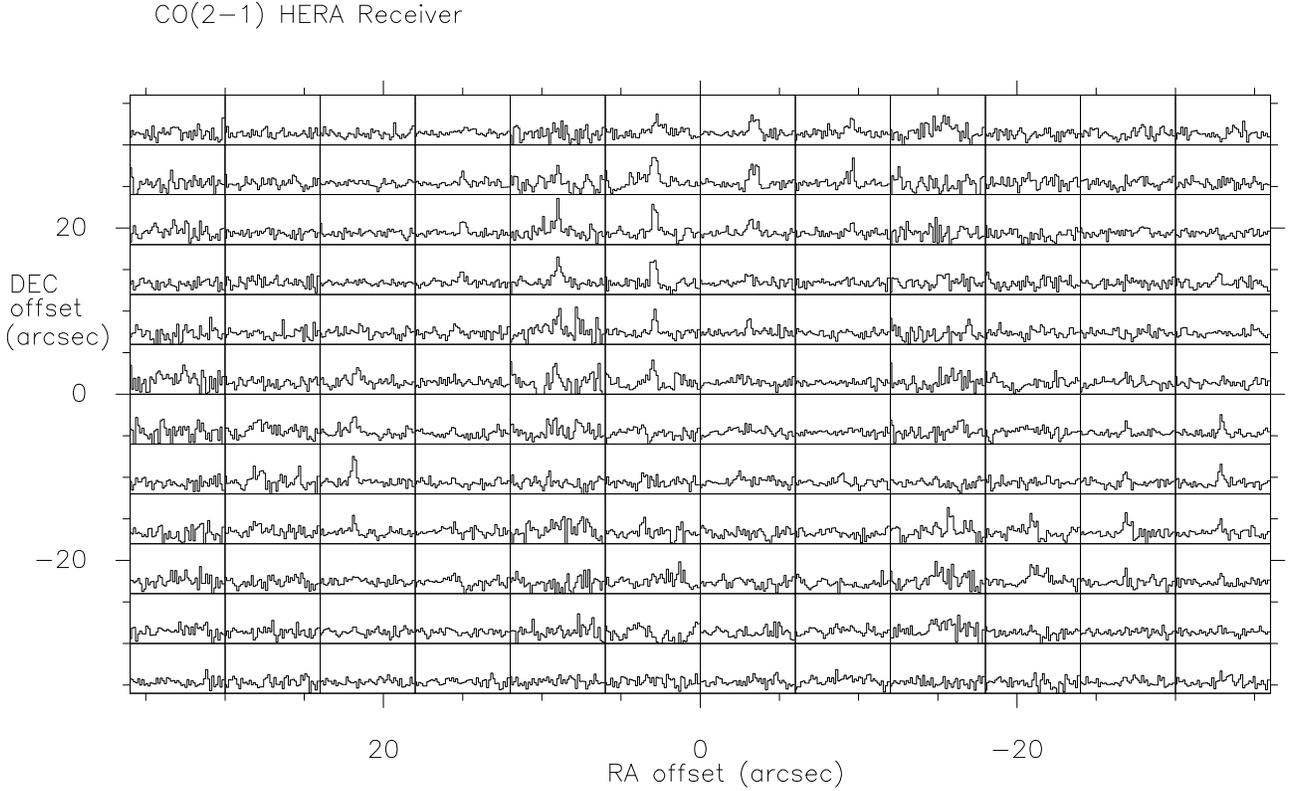}
    \caption{$^{12}$CO(2-1) map of the M~81 central region observed with the HERA receiver.
           Each spectrum has a velocity scale from -600 to 600 km s$^{-1}$ and a temperature scale from -20
           to 70 mK. The positions are offsets relative to the 
M~81 nucleus assumed of coordinates RA$_{J2000.0}$ = 09$^{h}$ 55$^{m}$ 33$^{s}$, 
DEC$_{J2000.0}$ = 69$^{\circ}$ 03$^{\prime}$ 55$^{\prime\prime}$.}
 \label{hera_spectra}
   \end{figure*}

\begin{figure}[ht]
   \centering
   \includegraphics[width=9cm]{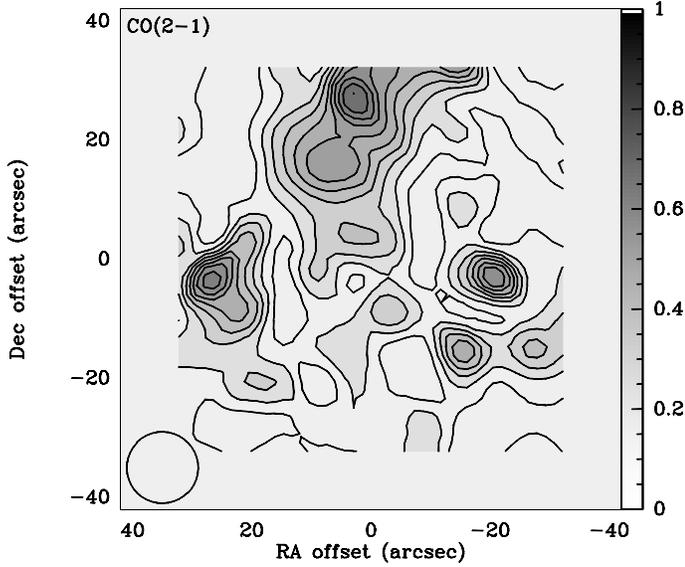}
   \caption{Integrated $^{12}$CO(2-1) emission. The contour levels are
            between 0.410 to 4.10 K km/s in steps of 0.410 K km/s. The beam of $12^{\prime\prime}$
            is indicated at the bottom left.}
  \label{fig3b}
  \end{figure}

\begin{figure}[ht]
   \centering
   \includegraphics[width=9cm]{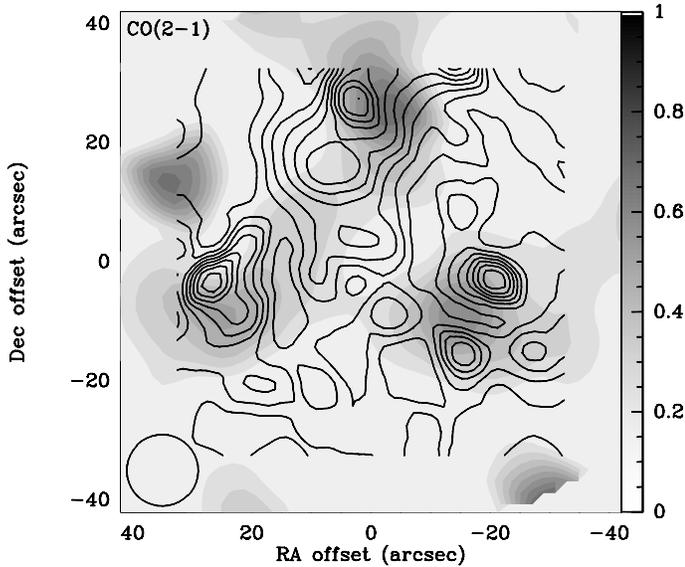}
   \caption{Integrated $^{12}$CO(2-1) emission detected with the HERA receiver superposed on
   the $^{12}$CO(2-1) emission found with A and B receivers. The contour levels are
   the same as shown in Fig. \ref{fig3b} between 0.410 to 4.10 K km/s in steps of 0.410 K km/s and the intensity levels
   detected with A and B receivers are in grey scale. The beam of $12^{\prime\prime}$
   is indicated at the bottom left. The intensity peaks detected with the two observing configurations
   agree quite well between them.}
   \label{hera-aeb}
   \end{figure}

  The observations done with the HERA receiver cover the central $\sim$ 1.3 kpc,
similar at the central $\sim$ 1.6 kpc reported above observed with the A and B receivers.
These observations confirm that the molecular gas emission in the M~81 center is at a low
temperature. Fig. \ref{hera_spectra} shows the spectral map and Fig. \ref{fig3b} the integrated contours
obtained with our observations.
HERA observations performed for the $^{12}$CO(2-1) line also reveal ``islands'' of molecular
gas in the nuclear region. There are regions completely devoid of emission, but also central offsets 
with a faint but detectable emission. The pointings with signals $\ge$ 2$\sigma$ have peak temperatures 
between 16 mK and 65 mK. There is good correspondence between the emission in CO(2-1) found with HERA 
receiver and with A and B receivers, as shown in Fig. \ref{hera-aeb} where the distributions found 
with the two observing configurations overlap.

\subsection{The CO(2-1) to CO(1-0) line ratio}
   The combination of observations of the CO(1-0) and CO(2-1) lines allow us to compute the line ratio
and to study the physical properties of the gas more in detail, such as excitation temperature and 
optical depth.

   The interpretation of the $R_{21}=I_{21}/I_{21}=\int{T_{mb}(2-1)}$dv /$\int{T_{mb}(1-0)}$dv line 
ratio is quite complex and usually simplified with \textit{standard} assumptions. The main 
assumption is to separate the analysis of the $R_{21}$ ratio into two limiting cases: the optically thin
case and the optically thick case. The analysis of these two limiting cases for the CO molecule has 
shown two important things. First, for high temperatures, the $R_{21}$ ratio asymptotically approaches a 
value of 4 in the optically thin case, while it only approaches 1 in the optically thick case. If 1 is 
the limit for an optically thick gas,  $R_{21}$ ratios $>$ 1 are signatures of optically thin CO emission. 
Second, in the optically thin limit, the $R_{21}$ ratio does not drop below 1 until the excitation 
temperature drops below 8 K. This situation is improbable, but not impossible, especially if the excitation
of the gas is subthermal.

   The $^{12}$CO lines are optically thick in most galaxies, including the Milky Way, but there are other 
galaxies where the problem is still open, like M 82 -the interacting companion of M~81- \citep[]{loiseau}
or NGC 3628 \citep[]{reuter}. When studying the line ratio for a large sample (118) of nearby spiral galaxies, 
\citet{braine1} saw 
that subthermal excitation or different gas filling factors are only required for explaining line ratios 
less than 0.7.

\begin{table}[ht]
	\centering
	\caption[]{$R_{21}$ line ratio for spiral arms of M~81 by \citet{brouillet2}.}
		\begin{tabular}{lrccc}
		\hline
		Field       & Offsets &$\int{T_{mb}(1-0)}$dv &$\int{T_{mb}(2-1)}$dv & $R_{21}$ \\
		& ($^{\prime\prime}$,$^{\prime\prime}$)&[K km s$^{-1}$]&[K km s$^{-1}$]&\\
		\hline
\textbf{N1}	&	(0,0)	    &	1.14        &        2.11        &        1.85              \\
\textbf{N5}	&	(0,0)	    &	1.52        &        0.70        &        0.46               \\
   	        &	(10,0)	    &	1.27        &        1.05        &        0.83               \\
            	&	(-10,0)	    &	1.39        &        0.70        &        0.50               \\
	        &	(20,0)	    &	0.89        &        0.53        &        0.59               \\
	        &	(-20,0)	    &	1.52        &        1.05        &        0.69               \\
	        &	(20,-10)     &	2.03        &        1.05        &        0.52               \\
\textbf{N7}	&	(0,0)	    &	2.53        &        1.23        &        0.49               \\
	        &	(20,0)	    &   1.52        &        1.93        &        1.27               \\
	        &	(-20,0)	    &	2.28        &        1.05        &        0.46               \\
	        &	(0,20)	    &	0.51        &        1.40        &        2.77               \\
	        &	(0,-20)	    &	1.39        &        4.04        &        2.90               \\
\textbf{S2}	&	(-10,0)	    &	0.76        &        1.05        &        1.39               \\
 	        &	(0,-10)	    &	2.03        &        2.28        &        1.13               \\
	        &	(20,0)	    &	2.03        &        1.40        &        0.69               \\
     		&	(-20,0)	    &   1.65        &        1.40        &        0.85               \\

	   \hline
		 \end{tabular}
		 \label{table3}
     \end{table}

With line ratios of 0.7, 0.8, and 0.9, the excitation temperatures are expected to be 
7 K, 10.5 K, and 21 K, respectively. These low line ratios can be explained considering 
that at high gas densities in the heating mechanism the dust-gas collisions are more important 
than cosmic rays that ionize H$_{2}$ or via formation of H$_{2}$ on dust grains. When 
neglecting the heating from gravitational collapse or magnetic sources and assuming the gas in local 
thermal equilibrium (T$_{ex}$ = T$_{kin}$), \citet{braine1} derived a relationship between gas temperature and dust 
temperature at high densities. At gas densities of 10$^{4}$ cm$^{-3}$, which is required to approximately 
thermalize the J = 2 level of CO, and dust temperatures of 30 K, the equilibrium gas temperature 
is 10 - 13  K. Considering that in the clouds the dust is probably warmer in the vicinity of 
OB stars  and lower elsewhere, the gas heated by the warm dust at 50 - 60 K will reach the 
thermal equilibrium at temperatures of 15 - 20 K. The J = 2 level of CO is populated at these 
temperatures, and \citet{braine1} computed that the expected line ratios are between 0.7 and 0.9.
In addition, it is important to remember that the CO emission is not always 
thermalized and that the excitation temperature may change along the line of sight and within a single beam.

   We computed the $R_{21}$ integrated intensity ratios for the offsets observed in the M~81 center,
after smoothing the $^{12}$CO(2-1) data to the $^{12}$CO(1-0) beam resolution 
($22^{\prime\prime}$) and correcting for the different beam efficiencies.
   In Table \ref{table2} the main results of the Gaussian fits of the observations are listed. 
   In this table, columns (1) and (2) are the offsets
   ($\Delta\alpha$ and $\Delta\beta$) referred to the center of the galaxy, which
   coordinates are given in Table \ref{table1}; columns (3) and (6) are the beam-average
   radiation temperatures ($T_{mb}$) for the two transitions; columns
   (4) and (7) are the line-widths ($\Delta$v) for the two transitions; column (9) is 
   the $R_{21}=I_{21}/I_{10}$ convolved line ratios; columns
   (5) and (8) are the CO intensities of the two transitions obtained from Gaussian fits.

The first striking note is the low temperature both in CO(1-0) and CO(2-1) transitions, which 
characterizes the emission of all observed positions in the M~81 central region. For those detections 
$\geq2\sigma$, the average brightness temperature in CO(1-0) is 33.44 mK, while in CO(2-1) it is 
21.47 mK.

   The $R_{21}$ ratios found in the central regions of M~81 are quite low: where the CO(1-0) emission 
is particularly weak, that of the CO(2-1) line appears weaker or difficult to detect, producing a low 
$R_{21}$ value. The average $R_{21}$ ratio for the center of M~81 is 0.68, a value atypical of 
galactic nuclei. Actually, in CO surveys \citep{braine1} the central regions (inner kpc) of the galaxies 
have, on average, $R_{21}\sim$0.89, rarely reaching higher ratios (e.g. for NGC 3310). According to 
\citet{braine1}, this happens because the inner molecular gas is optically thick in $^{12}$CO lines
and is cool but not cold (T $>$ 10 K). Our low ratio for M~81 is lower than the average value of 
0.89 for galaxies similar in morphology and distance. In addition, isolated galaxies appear to have 
line ratios that are lower than interacting/perturbed galaxies \citep{braine1}. M~81 exhibits a low value, 
despite being a classical example of interaction. 

Studies of the disk in single galaxies show, in general, lower line ratios  
than those in the galactic nuclei. IC 342 \citep{eckart}, NGC 6946 \citep{casoli}, and M 51 
\citep{garcia-burillo} are examples in which the $R_{21}$ ratio typically decreases from $\sim$1 
in the galactic nuclei to 0.4-0.6 in the spiral arms.

   To test if $R_{21}$ in M~81 shows a similar trend and the ratio decreases, which is necessary to use data in the literature. 
There is only one observational study on M~81 with data of the spiral arms in both CO(1-0) and CO(2-1) 
lines \citep{brouillet2}.  We used Fig. 2 of \citet{brouillet2} to study only the fields with detections 
for the two lines and computed the corresponding $R_{21}$ ratios after the transformation from
detected $T^{*}_{A}$ to $T_{mb}$, needed for the comparison with our data. The results of this 
analysis are collected in Table \ref{table3}. In this table, columns (1)
	and (2) are the fields and the corresponding coordinates respectively defined
	in Fig. 2 of \citet{brouillet2}; columns (3) and (4) are the intensities lines $I_{10}$
	and $I_{21}$ respectively computed from the $T^{*}_{A}$ detected by \citet{brouillet2};
	column (5) is the $R_{21}=I_{21}/I_{10}$ line ratio. 
	Table \ref{table3} shows that the $R_{21}$ line ratio spans in a wide 
 range of values, without a clear radial trend. For instance, the regions N5 and N7, that are 
 in the same area of the northeastern spiral arm, show 
 very different values for $R_{21}$. This confirms the ``anomalous'' $R_{21}$ behavior of M~81 with 
 respect to other spiral galaxies.

In conclusion, M~81 is a galaxy with low intensity and an anomalous behavior of CO lines. In this
respect, it is similar to the central CO-poor galaxies NGC 7331 \citep{young1,young3}, which has 
$R_{21}\sim0.54$ \citep{israel} and M~31, where similar low ratios have been found for individual 
dark clouds in the central region \citep{allen}. In this case, the authors attributed this fact 
to a subthermal excitation of the gas, at least for the CO(2-1) transition. 
These low ratios may be understood if coming from cool or cold, diffuse gas with a molecular hydrogen 
density (n(H$_{2}$)) not high enough to excite the CO molecules.

\begin{table*}[ht]
	 \centering
	 \caption[]{GMAs identified in the M~81 center with GAUSSCLUMPS.}
	 \begin{tabular}{lrrcccccccc}
	 \hline
	  Number & $\Delta\alpha$ & $\Delta\delta$ & v$_{0}$ & $T_{0}$ & $\Delta$x$_{1}$ & $\Delta$x$_{2}$ & $\varphi$ & $\Delta$v & dv/dr & $\varphi_{v}$\\
	  & [$^{\prime\prime}$] & [$^{\prime\prime}$] & [km s $^{-1}$] & [K]& [$^{\prime\prime}$] & [$^{\prime\prime}$] &[$^\circ$]& [km s $^{-1}$] & [km s $^{-1}$]&[$^\circ$] \\
	  (1)&(2)&(3)&(4)&(5)&(6)&(7)&(8)&(9)&(10)&(11)\\
	 \hline
1	&	6.9	&	20.0	&	6.13	&	0.07	&	15.8	&	19.2	&	83.7	&	17.2	&	1.67	&	88.2	\\
2	&	9.4	&	8.8	&	238.31	&	0.06	&	12.0	&	12.1	&	5.9	&	12.1	&	2.76	&	-111.8	\\
3	&	-14.8	&	-16.1	&	133.76	&	0.05	&	12.0	&	12.1	&	101.3	&	12.1	&	2.29	&	-106.3	\\
4	&	-7.1	&	27.3	&	132.12	&	0.05	&	12.0	&	17.2	&	94.5	&	13.9	&	2.05	&	-69.7	\\
5	&	4.4	&	-20.2	&	341.63	&	0.05	&	12.0	&	12.1	&	36.9	&	12.1	&	1.72	&	-138.2	\\
6	&	8.9	&	2.7	&	524.09	&	0.05	&	12.0	&	19.0	&	43.8	&	14.3	&	2.04	&	-90.2	\\
7	&	-17.3	&	-20.6	&	-21.47	&	0.04	&	12.0	&	17.2	&	48.5	&	13.9	&	1.40	&	-86.3	\\
8	&	9.2	&	-28.2	&	264.09	&	0.04	&	12.0	&	12.1	&	56.9	&	12.1	&	3.04	&	-149.6	\\
9	&	-15.7	&	-20.2	&	185.01	&	0.04	&	12.0	&	16.8	&	65.1	&	13.8	&	1.69	&	44.6	\\
10	&	-14.7	&	-27.0	&	263.61	&	0.04	&	12.0	&	12.1	&	63.6	&	12.1	&	4.41	&	28.6	\\
11	&	7.9	&	9.0	&	30.24	&	0.04	&	12.0	&	13.0	&	121.2	&	12.5	&	3.42	&	-33.7	\\
12	&	1.6	&	27.2	&	56.42	&	0.04	&	12.0	&	23.6	&	88.8	&	15.1	&	1.39	&	-105.3	\\
13	&	8.4	&	-15.5	&	316.24	&	0.04	&	12.0	&	12.1	&	87.7	&	12.1	&	5.83	&	4.7	\\
14	&	9.2	&	-14.9	&	82.50	&	0.04	&	12.0	&	12.1	&	22.6	&	12.1	&	3.62	&	115.8	\\
15	&	-32.9	&	-9.2	&	-22.05	&	0.04	&	12.9	&	20.0	&	81.0	&	15.3	&	1.04	&	-18.8	\\
16	&	-14.5	&	21.9	&	-22.41	&	0.04	&	19.0	&	12.1	&	29.3	&	14.5	&	2.18	&	42.0	\\
\hline
\label{table4}
\end{tabular}
\end{table*}

  \section{Clumping properties of the gas}
   There is evidence that molecular clouds have a high degree of internal fragmentation on all 
scales: this medium is defined as clumpy.
\citet{regan} and \citet{helfer}, used the BIMA SONG survey -a systematic imaging survey 
in CO(1-0) emission of a sample of 44 nearby spiral galaxies- to study the molecular gas emission 
in the centers and inner disks of galaxies in their sample with a resolution of $6^{\prime\prime}$. 
Their CO maps display a remarkable variety of molecular gas morphologies, and the CO surface brightness 
distribution shows substructures with dimensions of a few hundred parsecs. \citet{sakamoto1}
found a similar result
with a CO aperture-synthesis of the central regions of 20 nearby spiral galaxies using the
Nobeyama and Owens Valley millimetric arrays with a typical resolution of $4^{\prime\prime}$.
 In many galaxies they detected CO clumps of subkiloparsec size that they defined
as giant molecular associations (GMAs).

 In Figs. \ref{fig3b} and \ref{hera-aeb} it appears difficult to distinguish individual
structures, such as molecular clouds -probably present- and estimate their characteristic
parameters (dimension, mass, etc.). To study the molecular gas structure on a small scale, it is
necessary to use specific tools to identify single CO clouds.

There are several different methods of decomposing clumps in a 3-dimensional space;
i.e. position-position-velocity space, and all the techniques developed allow 
characterization of the clumpy structure: size, line width, 
and one of the most important parameters of a clump, the mass.
The algorithms generally used to study the clumpy nature of the gas are
GAUSSCLUMPS \citep{stutzki,kramer} and GAUSSFIND \citep{williams}. GAUSSCLUMPS
decomposes the 3-dimensional data cubes into a series of clumps, where each clump
corresponds to a peak position and is assumed to have a Gaussian shape.
GAUSSFIND works on the 3-dimensional data cubes as the eye would find a data set by using local
maxima. Usually, the results of the two algorithms agree with each other.
In general, GAUSSCLUMPS is able to more easily find low-mass clumps than
GAUSSFIND; and since M~81 shows a weak CO emission, we decided to use GAUSSCLUMPS.
In addition, this algorithm can work with the package GILDAS, already used here.

   \subsection{Identification of molecular associations}
  The first application of GAUSSCLUMPS to CO(2-1) data obtained using HERA receiver
  found 27 molecular associations in the central 1.3 kpc; however, it is 
  necessary to fix the constraints to check misidentification.
  The algorithm forces the unresolved 
  clumps to have a size corresponding to the spatial and velocity resolutions, so if the 3D size of 
  a clump is exactly equal to the resolution in each axis, this clump could be an artefact. With a beam
  of $12^{\prime\prime}$, only clumps with the intrinsic FHWM size $\Delta\xi >10^{\prime\prime}$ are 
  realistic, so all the other clumps have been discarded.
   The second constraint is on the intrinsic brightness temperature, which must be at least 5 times 
the noise level of the original map, T $>$ 0.035 K \citep{stutzki}.  After applying these constraints, 
the number of molecular associations found in the M~81 central region has become 16 and their properties are listed 
in Table \ref{table4}. In this table, column (1) is the GMA number; columns (2) and (3) are the GMA center positions
referred to the center of the galaxy, which coordinates are given in Table \ref{table1};
column (4) is the GMA center velocity;
column (5) is the GMA peak intensity;
column (6) and (7) are the GMA spatial FWHMs along the two principal axis;
column (8) is the orientation of the first principal axes;
column (9) is the GMA velocity FWHM;
column (10) is the GMA internal velocity gradient;
column (11) is the direction of internal velocity gradient.

 Considering that the Gaussian-shaped molecular associations have two principal axes, we find 
 a mean intrinsic diameter for all molecular structures identified of 14$^{\prime\prime}$ 
 ($\sim$250 pc). 
 Molecular structures with these dimensions are defined as GMAs.
 The maximum GMA peak intensity is 0.070 K, the mean intensity is 0.046 K 
 and the mean velocity FWHM 13.5 km s$^{-1}$.

   Summing up the area corresponding to the intrinsic FWHM extent of the fitted GMAs
   in the selected field, one obtains a total area covered by GMAs that is from 
   2.2 to 4.5 times smaller than the field, therefore the filling factor of the GMAs is low, about 0.3.

\begin{table*}[ht]
\centering
\caption[]{Properties of the GMAs derived by application of GAUSSCLUMPS.}
\begin{tabular}{ccccccccc}
\hline
Number & $R_{eff}$ & $M_{thin}$ & $M_{thick}$ &$M_{vir}$ & $\alpha_{vir}$ & $X$\\
&&$10^{3}$&$10^{5}$ &$10^{6}$ &&$10^{21}$\\
& [pc]& [$M_\odot$] &[$M_\odot$] &[$M_\odot$]& & [mol cm$^{-2}$ (K km s$^{-1}$)$^{-1}$]\\
(1)&(2)&(3)&(4)&(5)&(6)&(7)\\
\hline
1	&	151.88	&	3.83	&	7.20	&	8.54	&	11.86	&	2.29	\\
2	&	105.08	&	1.11	&	2.08	&	2.92	&	14.07	&	2.71	\\
3	&	105.08	&	0.92	&	1.73	&	2.92	&	16.89	&	3.26	\\
4	&	125.28	&	1.50	&	2.83	&	4.60	&	16.27	&	3.14	\\
5	&	105.08	&	0.92	&	1.73	&	2.92	&	16.89	&	3.26	\\
6	&	131.67	&	1.71	&	3.21	&	5.12	&	15.93	&	3.07	\\
7	&	125.28	&	1.20	&	2.26	&	4.60	&	20.34	&	3.92	\\
8	&	105.08	&	0.74	&	1.38	&	2.92	&	21.11	&	4.07	\\
9	&	123.81	&	1.17	&	2.19	&	4.48	&	20.43	&	3.94	\\
10	&	105.08	&	0.74	&	1.38	&	2.92	&	21.11	&	4.07	\\
11	&	108.91	&	0.82	&	1.54	&	3.23	&	21.04	&	4.06	\\
12	&	146.74	&	1.79	&	3.37	&	6.36	&	18.86	&	3.64	\\
13	&	105.08	&	0.74	&	1.38	&	2.92	&	21.11	&	4.07	\\
14	&	105.08	&	0.74	&	1.38	&	2.92	&	21.11	&	4.07	\\
15	&	140.06	&	1.66	&	3.11	&	6.23	&	20.03	&	3.86	\\
16	&	132.22	&	1.40	&	2.63	&	5.28	&	20.10	&	3.88	\\
\hline
\label{table5}
\end{tabular}
\end{table*}

 \subsection{GMAs masses}
 For an optically thin gas, or at the other extreme for a completely optically thick gas \citep{kramer}, 
the mass of a molecular association is proportional to the fitting parameters of GAUSSCLUMPS: the 
molecular association intensity, the molecular association velocity width, and the spatial 
FWHMs along both two principal axes:
\begin{eqnarray}
      M \propto T_{pk}\times \Delta {\rm v} \times \Delta x_{1}\times \Delta x_{2}.
  \label{mass1}
  \end{eqnarray}
  
We now present mass computations in the two hypotheses below.
For the estimation in the optically thin case, we adopt the Rayleigh-Jeans approximation,
where: $ f(T_{ex}) = \frac{h \nu /k }{1-exp\left[-h \nu/kT_{ex}\right]} \sim T_{ex} $,
since $T_{ex} >$ 15 K. LTE is also assumed, meaning that $T_{ex}$ is the same for all levels.  
The total CO column density $N_{CO}$ is related to the population $N_i$ of level $i$ by 
$ N_i/N_{CO} = g_i exp\left[-E_i/kT_{ex}\right]/Q$, where $g_i$ and $E_i$
are the degeneracy factor and the energy of the level $i$, and $Q$ the partition 
function is $Q(T) = 2 kT / h \nu_{1-0} = T (K) /2.76$.
This total CO column density, derived from the integrated intensity of the CO(2-1) line,
is given by the following equation, in the optically thin limit:
\begin{eqnarray}
     \frac {N_{CO}}{\left(mol/cm^{2}\right)}=1.2\times 10^{13} \times T_{ex}\left(K\right)\int T_{mb}\left(K\right){\rm dv}\left(km/s\right).
        \label{density2}
  \end{eqnarray}
The corresponding molecular mass (taking into account helium) is
\begin{eqnarray}
      \frac{M}{\left(M_\odot\right)}&=&1.2 \times 10^{13}\frac{m_{H_2}}{\left(M_\odot\right)}\frac{N(H_2)}{N(^{12}CO)}\times T_{ex}\left(K\right)\nonumber\\
	& & \times \int \int T_{mb}\left(K\right){\rm dv}\left(km/s\right)dA(cm^{2}).
	 \label{mass2}
  \end{eqnarray}
In terms of fitting parameters in Table \ref{table4} \citep[see][]{chi} and when assuming an excitation 
temperature of $T_{ex}\simeq15$ K, an abundance ratio $N(H_2)/N(^{12}CO)\simeq10^4$, and a distance 
of 3.6 Mpc, Eq. (\ref{mass2}), valid in the optically thin case, becomes

\begin{eqnarray}
	\frac{M_{thin}}{\left(M_\odot\right)}&=&\frac{T_{mb}}{\left(K\right)}\times \frac{\Delta {\rm v}}{\left(km/s\right)}\times \frac{\Delta x_{1}}{\left(^{\prime\prime}\right)}\times \frac{\Delta x_{2}}{\left(^{\prime\prime}\right)}\times10.5.
	\label{mass3}
\end{eqnarray}
 In this hypothesis, the ratio between the observed peak brightness temperature and the excitation 
temperature is the filling factor of the porous molecular associations, on the order of 3 $\times$10$^{-3}$.
 Applying Eq. (\ref{mass3}), we find very small molecular association masses contained in a low mass range,
 from $7.37\times10^{2} M_\odot$ to $3.83\times10^{3} M_\odot$, with a mean value of $1.31\times10^{3} 
M_\odot$. The masses computed for all molecular associations are listed in Table \ref{table5}.
In this table, column (1) is the GMA number as defined in Table \ref{table4}; column (2) is the effective radius of the GMAs; 
columns (3), (4), and (5) are the masses of the GMAs computed in the optically thin limit, in the optically thick limit, and using the 
virial theorem, respectively; column (6) is the ratio $\alpha_{vir}=M_{vir}/M_{obs}$ for each GMA, where $M_{obs}=M_{thick}$; 
column (7) is the derived $X$ conversion factor for each GMA.

 In the optically thick case, the molecular mass of the molecular associations is determined by the $H_{2}$ density 
($N_{H_{2}}$) using the \textit{standard} $X$ conversion factor value \citep[$X=2.3\times 10^{20}$ mol 
cm$^{-2}$ (K km s$^{-1}$)$^{-1}$,][]{strong}, $N_{H_{2}}=X \int T_{CO(1-0)}$dv. Since we resolve 
single GMAs in the CO(2-1) transition, using the mean $R_{21}=I_{21}/I_{10}$ value computed here
($\overline{R}_{21}=0.68$), the $H_{2}$ density can be expressed by

\begin{eqnarray}
	\frac{N_{H_{2}}}{\left(mol/cm^{2}\right)}&=&\frac{X} {\left(\frac{mol/cm^{2}}{K km/s}
	\right)} \times \frac{1}{R_{21}}
	\int T_{CO(2-1)}\left(K\right){\rm dv}\left(km/s\right) \nonumber\\
	&=&  3.4\times10^{20}\int T_{CO(2-1)}\left(K\right){\rm dv}\left(km/s\right),
	\label{density3}
\end{eqnarray}
and the corresponding molecular mass is
\begin{eqnarray}
	\frac{M}{\left(M_\odot\right)}&=& 3.4 \times 10^{20}\frac{m_{H_2}}{\left(M_\odot\right)}\nonumber\\
	&&\times \int \int T_{CO(2-1)}\left(K\right){\rm dv}\left(km/s\right)dA(cm^{2}).
	\label{mass4}
\end{eqnarray}
In terms of fitting parameters in Table \ref{table4} and assuming a distance of 3.6 Mpc, Eq. 
(\ref{mass4}) gives the GMA mass in the optically thick limit:
\begin{eqnarray}
	\frac{M_{thick}}{\left(M_\odot\right)}&=&\frac{T_{mb}}{\left(K\right)}\times \frac{\Delta {\rm v}}{\left(km/s\right)}\times \frac{\Delta x_{1}}{\left(^{\prime\prime}\right)}\times \frac{\Delta x_{2}}{\left(^{\prime\prime}\right)}\times 1.97 \times 10^{3}.
	\label{mass5}
\end{eqnarray}
Applying Eq. (\ref{mass5}), the GMA masses range from $1.38\times10^{5} M_\odot$ to 
$7.20\times10^{5} M_\odot$, with a mean value of $2.46\times10^{5} M_\odot$. The total mass of the 
GMAs is $3.22\times10^{5} M_\odot$. The masses computed for all GMAs are listed in Table \ref{table5}.
 
 That the masses obtained through the optical thin hypothesis are much lower than those 
calculated in the optically thick case is not ``anomalous'': instead, it is expected that the optical thin 
hypothesis underestimates the masses, since GMAs have a rich substructure of 
subclumps with high column density. The subclumps are probably optically thick, as is generally the case in external 
galaxies, as suggested by the high $^{13}$CO emission commonly observed. Anyway we decided to 
also analyze the optically thin case because M~81 represents an extraordinary laboratory where 
all possibilities should be considered even if unusual for similar galaxies.

 \subsection{Virial equilibrium}
 An alternative way of determining GMA masses is to compute their virial mass. 
 Assuming a balance between the total kinetic energy $T$ of a GMA and the gravitational potential 
$\Omega$, and neglecting external pressure, the virial mass $M_{vir}$ of the clouds can be calculated 
using the formula

\begin{eqnarray}
	M_{vir}=K\times D_{e} \times \Delta {\rm v}^{2}
	\label{mass6}
\end{eqnarray}
where $M_{vir}$ is in $M_\odot$, $D_{e}$ is the effective diameter in pc of the equivalent sphere of 
radius $R=(\Delta x_{1}\Delta x_{2}/4)^{0.5}$, and $\Delta$v is in km s$^{-1}$ \citep{solomon2}. 
We assume that the cloud is a sphere with a power-law ($\alpha=1$) density distribution. In this case,
the constant K in formula (\ref{mass6}) is equal to 95. The rougher assumption of uniform density 
(\ref{mass6}) would have instead produced K=104. Here, the hypothesis of spherical shape for the cloud 
makes the formula independent of the galaxy inclination.

 A comparison between the virial mass and the mass computed from the observed intensity lines 
($M_{obs}$) in the optical thick hypothesis is informative for the physical state of the clouds, 
their optical thickness, and for the actual X= N(H$_2$)/I(CO) conversion factor.

 The virial masses of the 16 GMAs we identified span from $2.92\times 10^{6}$ to $8.54\times 10^{6}
M_\odot$, with a mean value of $4.31\times10^{6} M_\odot$. 
In calculating $M_{vir}$, we
have neglected the mass of the atomic hydrogen ($M_{HI}$ = 0) since \citet{allen2}, studying the HI profile in M~81, 
found a deficiency of atomic hydrogen in the central $\sim$3 kpc, and the HI emission only begins to be significant 
in the inner spiral arms beyond $\sim$5 kpc from the nucleus.

We also checked that the stars do not contribute significantly to the mass
inside the volume of a GMA. In the hypothesis that the H$_2$ mass is computed from
the virial mass, resolved GMAs have mean surface densities ranging from 84 M$_\odot$/pc$^{2}$ to 118 
M$_\odot$/pc$^{2}$ with a mean value of 91 M$_\odot$/pc$^{2}$. The mean value of the 
mass-to-light ratio of visible matter 
expected for M~81 is M/L$_B$ = 5.5 M$_\odot$/L$_\odot$ \citep[see][]{tenjes}.
From the exponential surface brightness distribution of the stellar disk of the 
galaxy, we then deduced a mean stellar surface density 
of 36 M$_\odot$/pc$^{2}$ in the region. In addition, if we consider that GMAs normally 
have a thickness of about 50 pc while the stars are distributed in regions that are at least 
ten times thicker, the actual mean volumic stellar density 
is negligible with respect to the molecular density inside a GMA.

 All the GMAs have a virial mass
$\gg$ of the mass computed from CO observations, and the mean ratio $\alpha_{vir}=M_{vir}/M_{obs}$
computed for our GMAs is $\alpha_{vir} \simeq (10-20)\gg1$. The $\alpha_{vir}$ ratios computed for
all the GMAs are listed in Table \ref{table5}.

 \subsection{The $N(H_{2})/I_{CO}$ conversion factor}
 The value of the conversion factor between the $H_{2}$ column density of the molecular hydrogen
$N(H_{2})$ and the CO integrated intensity ($X=N(H_{2})/I_{CO}$) is still subject to debate.
Usually the \textit{standard} value of X found for the Milky Way is universally adopted for all
the galaxies and for all the regions of the same galaxy \citep[$X=2.3\times 10^{20}$ mol cm$^{-2}$
(K km s$^{-1}$)$^{-1}$,][]{strong}. This assumption often turned out not to be completely correct.

 The $X$ conversion factor is determined by various factors, such as the metallicity, the temperature,
the cosmic ray density, and the UV radiation field \citep[see][]{maloney,boselli}. For instance,
galaxies with lower metallicity generally have a higher $X$ factor; when studying 5 local
group galaxies, \citet{wilson3} found that the conversion factor increases by a factor of 4.6 when metallicity 
decreases of a factor 10. The $X$ value also changes with the morphological type of the galaxies:
galaxies earlier than Scd show values comparable to, or lower than, the Galactic one, while extremely
late-type spirals or irregular galaxies tend to show higher values \citep[][]{nakai}. A dependency of 
the $X$ conversion factor on $I_{CO}$ is also suspected: galaxies or observed positions with low 
$I_{CO}$ ($I_{CO}<$ 20-40 K km s$^{-1}$) show a higher $X$. An example of this behavior is the galaxy 
M 51, where $I_{CO}<$ 20 K km s$^{-1}$ and the $X$ derived value is a factor 2.6 lower than the mean 
Galactic value \citep[see Fig. 3 of][]{nakai}.

\begin{table*}
   \caption[]{Comparison of the results obtained applying the aperture photometry at the FUV-GALEX image of M~81
   in different regions observed in CO(1-0).}
   \begin{center}
   \begin{tabular}{lcccc}
   \hline
   \hline
   Reference             &  I$_{CO}$       & F$_{FUV}$& $G_{0}$ & I$_{CO}$/F$_{FUV}$ \\
                         &                 & 10$^{-14}$&& 10$^{14}$  \\
                         & [K km s$^{-1}$]   & [erg cm$^{-2}$ s$^{-1}$ $\AA^{-1}$] && [K km s$^{-1}$/(erg cm$^{-2}$ s$^{-1}$ $\AA^{-1}$)]\\
   \hline
   \citet{brouillet3}    & 25.90           & 1.37               &0.51              & 18.88\\
   \citet{knapen}        & 20.26           & 1.48               &0.35              & 13.65 \\
   This paper            & 53.50            & 1.13              &0.31               & 47.18 \\
   \hline
   \end{tabular}
   \label{table6}
   \end{center}
   \end{table*}

 Since M~81 shows atypical properties for its molecular gas, the CO to 
H$_2$ conversion ratio is also suspected of departing significantly from the mean Galactic value and in 
general from that of galaxies with the same morphological type. Thus, M~81 represents an exceptional 
laboratory to explore the variation of the $X$-ratio.

 We derived the $X$ conversion factor for all individually resolved molecular associations assuming the virial 
 equilibrium for single molecular complexes and using the large velocity gradient (LVG) 
 approximation for the formation of the CO line \citep{young2,sanders}. 
 This method has also been used to derive the conversion factor in nearby spiral galaxies such as M~31, M~33, 
 and the Magellanic Clouds \citep{wilson1,wilson2,arimoto}.

 In practice, we computed the X ratio (=$N(H_{2})/I_{CO}$) deriving the $H_{2}$ density from the virial masses 
 ($N(H_{2})=M_{vir}/(m_{H_2} \times dA)$, where $m_{H_2}$ is the mass of the hydrogen molecule and $dA$ 
 the area occupied by the cloud),
 and $I_{CO}$ is the CO observed intensity. For the individually resolved GMAs of M~81, assumed virialized,
 we found that the $X$ conversion factor takes a 
 much higher value than the mean Galactic value for all clouds, ranging from $\sim$10 to $\sim$18 
 times higher than the \textit{standard} $X$ and a mean value of $X=3.6\times10^{21}$ mol cm$^{-2}$ 
 (K km s$^{-1}$)$^{-1}$. The $X$ conversion factor values computed for all the GMAs are listed in 
 Table \ref{table5}.
 
 Our $X$ factor values are high, not only if compared to the standard Galactic factor, but also 
 with respect to values found in other nearby galaxies where the X ratio has been obtained with the 
 same method we used. Studying a sample of nearby spiral and dwarf irregular galaxies, \citet{arimoto}
 derived $X$ values using CO data found in the literature and computed the corresponding virial mass. 
 They found galaxies with both $X$ values lower and higher than in the Galaxy, such as M~31 with 
 $X=1.8\times10^{20}$ mol cm$^{-2}$ (K km s$^{-1}$)$^{-1}$, M~33 with $X=4.1\times10^{20}$ mol cm$^{-2}$ (K km s$^{-1}$)$^{-1}$, 
 and IC~10 where $X=6.6\times10^{20}$ mol cm$^{-2}$ (K km s$^{-1}$)$^{-1}$. 
 Even if they found high X values, these $H_{2}$-to-CO conversion factors are smaller than our X ratios 
 obtained for M~81. \citet{arimoto} found a $X$ ratio (=$3.1\times10^{21}$ mol cm$^{-2}$ 
 (K km s$^{-1}$)$^{-1}$) only for the SMC within the same order of magnitude of M~81. They specify 
 that this value has been obtained 
 by considering virial masses and CO luminosities for larger-scale complexes in the SMC
 and that this high $X$ would be due to a smaller amount of CO in a diffuse medium that
 is metal dependent, as well as photo-dissociated by the strong UV field, and can be present 
 only in dense molecular clouds.

 \section{Discussion}
 
   \subsection{Comparison with previous works}
   Our molecular gas detections in the center of M~81 agree with the only two works present in the 
literature that report CO emission in the nuclear region of this galaxy \citep{sage,sakamoto2}.
In comparison with the giant molecular clouds (GMCs) in the outer regions of M~81 
\citep{taylor2,brouillet3}, the CO emission that we detect in the M~81 center is particularly cold and 
unusual.
   The masses we find for GMAs in the CO(2-1) map are similar to the higher masses observed in M 33 
\citep{wilson1}, and higher than what is measured in the SMC \citep{rubio}.

   Our first result is that the nuclear region of M~81 (within $\sim$300 pc) appears devoid of CO 
emission, in agreement with \citet{brouillet1} and \citet{sakamoto2}. However, we detect CO emission
outside the 300 pc radius. This supports the claims of \citet{sage} who, found CO emission when observing 
one offset towards the 
center galaxy with the NRAO 12m telescope and a beam of $55^{\prime\prime}$ \citep[see 
Fig. 1 of][]{sage}. There is a good correspondence between our velocity profile and theirs.

   A more useful and interesting comparison can be done between our results and those of 
\citet{sakamoto2}.  In detail, the ``pseudoring'' that we detected at $\sim460$ pc from the nucleus in 
the northeast direction (Fig. \ref{fig2}) corresponds to the maximum intensity of \citet{sakamoto2}. 
Observing the central kiloparsec with the Nobeyama Millimetric Array-NMA (resolution of 
6.9 $^{\prime\prime}\times$ 5.8 $^{\prime\prime}$ that corresponds to $\sim$ 120$\times$100 pc at
the distance of M~81), they produced maps only for the $^{12}$CO(1-0) line.  Since they made 
interferometric measurements, with an FWHM primary beam of 1$^{\prime}$ and a beam of 
$\sim7^{\prime\prime}$, extended emission seen with the 22$^{\prime\prime}$ IRAM CO(1-0) beam is 
filtered out.

   In spite of this, the total mass of molecular gas that we find, assuming the same 
\textit{standard} X-factor value of \citet{strong}, is quite comparable to that of \citet{sakamoto2}: 
our resolved GMAs in CO(2-1) emission have a total mass of $3.22 \times 10^{6}M_\odot$, only a 
little bit lower than $\sim$ 1 $\times 10^{7}M_\odot$ found by \citet{sakamoto2} for the 
``pseudoring'' in CO(1-0) emission.
   In addition to the result of \citet{sakamoto2}, who found the molecular gas mainly located in a 
lopsided distribution in the northeast direction, we also detected some emission towards the 
southwest.
   Our peak temperatures for the $^{12}$CO(1-0) line are very low, ranging from $\sim$15 to $\sim$64 mK,
lower than the average peak brightness temperature ($\sim$0.3 K) of each GMA found by \citet{sakamoto2}.
This is compatible with the different beam sizes, the synthesized beam of the interferometer being 10 
times smaller in surface.

 Another useful comparison can be made with molecular clouds identifications 
 in other external galaxies. The masses ($ \sim 10^{5}M_\odot$) and the 
 diameters ($ \sim 250$ pc) of the CO clouds identified in 
 M~81 are not much unusual for giant galaxies of the same type \citep{taylor2}. Even if in 
general the GMCs and GMAs in the galaxies of the Local Group are very similar, there is evidence of 
variations in the properties of the molecular clouds of galaxies of different morphologies, which range 
from early-type spirals (M~31 and M~81) to late-type spirals (M 33 of Scd morphological type) and from 
dwarf irregulars (IC 10) to dwarf ellipticals (NGC 205). For the variety of characteristics found in 
clouds of galaxies of these sub-morphological types, \citet{taylor2} conclude that in general 
very late-type galaxies should have smaller and less massive GMCs than early-type galaxies. 
Our results agree with this conclusion.

Finally, we know that, in the Local Group and in systems in interaction with the Local Group, macroscopic 
properties of molecular clouds are related by power relationships, first noted by \citet{larson} 
and then studied in detail by \citet{solomon2}. In particular,  \citet{solomon2} found a tight 
relationship between the Galactic dynamical molecular cloud masses, measured by the virial theorem, and the CO 
luminosity ($M_{vir} \propto (L_{CO})^{0.81}$). For several external galaxies, this power law is even valid  
if with little variations of the power index \citep{rand1,rand2,rosolowsky}. It is interesting to test if 
M~81 also shows non-typical behavior in this case. Our very weak CO detections correspond to CO 
luminosities of the order of magnitude 10$^{4}$ K km s$^{-1}$ pc$^{2}$ with virial masses within an
order of magnitude of 10$^{6}$ M$_\odot$ (see Table \ref{table5}). Comparing our results with the 
Fig. 2 of \citet{solomon2}, we can see that with the $L_{CO}$ = 10$^{4}$ K km s$^{-1}$ pc$^{2}$ virial 
masses we infer are overestimated by more than one order of magnitude with respect to the typical trend 
in our Galaxy, but also in external galaxies. This result agrees with the indications given 
by the $\alpha_{vir}$ ratios calculated in Sect. 4.3, and in general it confirms the unusual 
scenario in the molecular gas properties of M~81.     
 
 \subsection{The $X$-factor problem}
 
 The $X$-factor peculiarity does not appear to be due to a particular metallicity, since M~81 has a 
 metallicity of Z=0.03, without any evident gradient from the central region to the outer disk \citep{kong}. 
 This ``normal'' metallicity value would produce a $X$ conversion factor similar to that of the Milky 
 Way. Some regions in M~81, however, have higher metallicity and are mostly located in the spiral 
 arms and HII regions. But in general, our estimate of the $X$-factor appears significantly high.

 A possible source of error may be the assumption of a fully virial equilibrium for the observed
 clouds. Under this hypothesis, we calculated the $X$-factor using $X=M_{vir}/L_{CO}$, where the 
 $M_{vir}$ has been computed with Eq. (\ref{mass6}), and $L_{CO}$ is the CO luminosity derived 
 from our observations. Application of the virial theorem to complex molecular structures may not 
 be completely appropriate, even if the expected deviation cannot be higher than a factor $\sim$2. 
 However, if the GMAs are not virialized and the \textit{standard} $X$ factor is applied, the 
 molecular content of M~81 would be exceptionally low, so M~81 would be peculiar in both cases.

 \subsection{Heating of the gas}
 The reason for the non-typical results on the molecular gas in M~81 is probably the 
exceptionally low CO excitation temperature that seems to characterize the center of this galaxy. 
Why is the molecular gas absent or subthermally excited in the M~81 center?

 A possible link could exist between low CO emission and the lack of ultraviolet emission that appears 
to characterize the M~81 central region. The ultraviolet image of M~81 realized with the satellite 
Galaxy Evolution Explorer (GALEX) clearly shows a lack of recent star formation, traced by the UV 
emission, in the nuclear region that we observed in CO.

 The existence of a link between molecular gas and far-ultraviolet (FUV) emission 
can be justified considering that the interstellar medium is excited, dissociated, and 
ionized by the FUV photons produced by young O and B stars. Atomic gas in the ISM recombines 
into molecular form mainly through the catalytic action of dust grain surfaces. Hydrogen 
nuclei cycle repeatedly from the molecular ($H_{2}$) to the atomic phase and back again at 
rates that depend on the incident FUV flux, the total volume density of the gas, and the 
dust-to-gas ratio. The action of the FUV flux that interacts with the surface of the 
molecular clouds can strongly affect their physics and chemistry by heating, dissociating, 
and exciting the gas.

In regions of the ISM where the physics is dominated by FUV photons -called photodissociation 
regions (PDRs)- useful information can be obtained on the physical state of the gas from observations 
of the spectral lines emitted by the excited atoms and molecules in those regions.

 In M~81 the possibility that low CO emission and weak FUV flux are correlated 
 has already been proposed by \citet{knapen}. When observing a region of the western spiral 
 arm in M~81, they explained the lack of or very weak 
CO(1-0) emission, observed far from FUV sources, as a consequence of insufficient
heating and excitation of the molecular gas. Usually, a lack in the CO emission is instead interpreted 
as due to the absence of measurable quantities of molecular gas or due to a low density gas not heated 
sufficiently to detect it. The limit of this interpretation, based on the linear proportionality between 
the CO surface brightness and the column density of the molecular hydrogen, is to neglect effects linked 
to local heating mechanisms. If the explanation of \citet{knapen} is correct, the flux of cosmic rays 
and the FUV surface brightness would be too low to heat the ISM, and in particular, the molecular component.
It is possible that this correlation of low-FUV/low-CO is also applicable to the central region that we 
observed.

 To verify this interpretation, we performed aperture photometry in the FUV emission on three fields 
already observed in CO: our central region of 1.6 kpc, the field of spiral arms observed by \citet{brouillet3}, 
and that - again on the spiral arms - observed by \citet{knapen}. The fields selected on the spiral arms 
are the same of Fig. 2 of \citet [][]{knapen}. We used the FUV-GALEX image of M~81 expressed in 
intensity and already sky-subtracted and performed the aperture photometry using the task QPHOT of 
IRAF, able to do a quick aperture photometry.
 This procedure allowed us to know the FUV flux emitted in regions observed in CO(1-0) and hence to 
compute the ratio between the two emissions in different fields, but of similar dimensions, of the 
galaxy. The aperture photometry was done with a radius of $40^{\prime\prime}$ for our central 
region and for the one observed by \citet{knapen} with the Nobeyama Radio Observatory (NRO), and 
with a radius of 
$32^{\prime\prime}$ for the field observed by \citet{brouillet3}. The results 
of the aperture photometry are summarized in Table \ref{table6}. 
In this table, column (1) gives the references relatively to CO(1-0) emission; column (2) is
  the CO(1-0) integrated intensity; column (3) is the FUV flux obtained by the aperture photometry;
   column (4) is FUV flux expressed in terms of the $G_{0}$ Hading unit;
   column (5) is the ratio between the CO(1-0) integrated intensity and the FUV flux.
We express the FUV fluxes in terms 
of the Hading unit, since it is one of the units usually used to model PDRs \citep[e.g.][]{kaufman2,allen3}.
In these models each one-dimensional semi-infinite slab of gas with constant density of H nuclei
is subjected to an equivalent one-dimensional flux of FUV photons $G_{0}$ measured in units of 
1.6 $\times$ 10$^{-3}$ ergs cm$^{-2}$ s$^{-1}$ over the photon energy range 6 - 13.6 eV. In these units, 
the average interstellar radiation field over 4$\pi$ sr is $G_{0}$ = 1.7 \citep[][]{draine,allen3}.

  The central region emits a weak FUV flux, within the same order of magnitude as that emitted in the two 
fields of the spiral arms, and the higher ratio between CO and FUV emissions in nuclear region
is mainly due more to the higher molecular gas intensity. If the low FUV emission is 
not able to heat the molecular gas sufficiently in the spiral arms, in the central region at the same quantity of FUV flux 
corresponds a molecular gas distributed in substructures, but subexcited. Probably the central region 
shows a higher CO/FUV ratio because in the spiral arms the molecular gas is less excited than the nucleus.

 \section{Conclusions}

 The physics of the molecular gas suggested by the observations reported in this paper shows M~81 
to be a galaxy that is not only CO-poor in its central region, but that also has molecular clouds with unusual 
properties, if compared to galaxies with similar distances and morphological type.
 The absent or very weak molecular gas emission in the nuclear region, the low $R_{21}$ line ratio, and 
a particularly high $X$ conversion factor value are its main peculiarities.
 The $N(H_{2})/I_{CO}$ conversion factor has a mean value of $X=3.6\times10^{21}$ mol cm$^{-2}$
(K km s$^{-1}$)$^{-1}$, 15.6 times higher than the $X$ value of \citet{strong} derived for our Galaxy.

 All unusual results found for the molecular component in the M~81 center are probably due to an 
excitation process of the gas non-typical for giant spirals in the Local Universe.
 The spiral arms of this galaxy are CO-poor, and it has been suggested that the FUV surface brightness 
is too low to heat the molecular gas component. The low CO emission we found in the center suggests 
that the gas in the nucleus is also subexcited, with even weaker FUV emission. We conclude that the 
lack of excitation of the gas, more than the absence of molecular gas, is the cause of the low CO 
emission there.

\begin{acknowledgements}
The observations reported here were made using the travel funds of RadioNet Networking Activities.
V. Casasola is pleased to acknowledge the hospitality and stimulating environment provided by the
Observatoire de Paris-LERMA, where part of the work on this paper was done during her stay in Paris,
thanks to Grant Vinci and the EARA agreement. The authors would like to thank the anonymous referee, whose
comments have been useful for improving the original version of the paper.

\end{acknowledgements}

\end{document}